\documentclass[aps,prb,twocolumn,superscriptaddress,color,pdflatex]{revtex4}

\usepackage{graphicx}
\usepackage{dcolumn}
\usepackage{amsmath,amssymb,bbold,bm,epsfig}

\begin{document}

\newcommand{\bk}{{\bf k}}
\newcommand{\bp}{{\bf p}}
\newcommand{\bv}{{\bf v}}
\newcommand{\bq}{{\bf q}}
\newcommand{\bs}{{\bf s}}
\newcommand{\bmm}{{\bf m}}
\newcommand{\tbq}{\tilde{\bf q}}
\newcommand{\tq}{\tilde{q}}
\newcommand{\bQ}{{\bf Q}}
\newcommand{\br}{{\bf r}}
\newcommand{\bR}{{\bf R}}
\newcommand{\bB}{{\bf B}}
\newcommand{\bE}{{\bf E}}
\newcommand{\bA}{{\bf A}}
\newcommand{\bK}{{\bf K}}
\newcommand{\vd}{{v_\Delta}}
\newcommand{\tr}{{\rm Tr}}
\newcommand{\bj}{{\bf j}}
\newcommand{\bn}{{\bf \hat{n}}}
\newcommand{\cH}{{\cal H}}
\newcommand{\cT}{{\cal T}}
\newcommand{\cM}{{\cal M}}
\newcommand{\BGamma}{{\bm \Gamma}}
\newcommand{\bsig}{{\bm \sigma}}
\newcommand{\bpi}{{\bm \pi}}

\title{Majorana Fermions in Proximity-coupled Topological Insulator Nanowires}

\author{A. Cook and M. Franz}
\affiliation{Department of Physics and Astronomy,
University of British Columbia, Vancouver, BC, Canada V6T 1Z1}

\begin{abstract}
A finite-length topological insulator nanowire, proximity-coupled to an ordinary
bulk $s$-wave superconductor and subject to a longitudinal applied
magnetic field, is shown to realize a one-dimensional topological
superconductor with unpaired Majorana fermions localized at both
ends. This situation occurs under a wide range of conditions and
constitutes what is possibly the most easily accessible physical
realization of the elusive Majorana particle in a solid-state system.

\end{abstract}
\maketitle
Predicted in a seminal 1937 paper \cite{majorana} as purely real 
(as opposed to complex-valued) 
solutions of the Dirac equation describing
a spin-${1\over 2}$ particle, Majorana fermions are distinguished by
the curious fact that they are their own anti-particles. More
precisely, in second quantized formulation, the creation and
annihilation operators for Majorana fermions coincide. In high-energy
physics compelling theoretical arguments suggest that neutrinos might
be Majorana fermions but a convincing experimental proof is yet to be
given \cite{wilczek1}. In condensed matter physics Majorana fermions
can appear as {\em emergent} degrees of freedom in certain systems of
electrons when superconducting order or strong correlations are
present \cite{franz1,stern1}. Over the past decade solid state
realizations of Majorana fermions have been under intense theoretical
study both as a fundamental intellectual challenge and as a possible
platform for fault-tolerant quantum computation
\cite{kitaev2,nayak1}. Yet their remarkable properties -- and indeed
their very existence -- await experimental confirmation.

The purpose of this Communication is to advance a proposal for a new type of
solid-state device that can serve as a host for Majorana fermions
under a wide range of experimentally accessible conditions. Our
proposed device, depicted schematically in Fig. 1, draws conceptually
on recent ideas to realize Majoranas in both two- and one-dimensional
heterostructures composed of a topological insulator \cite{fu2} or a
semiconductor with strong spin-orbit coupling
\cite{sau1,alicea1,sau2,oreg1}, coupled to an ordinary superconductor
through a proximity effect. Specifically, we consider a nanowire
(i.e. quantum wire with nanometer-scale cross-section) fashioned out
of a strong topological insulator (STI), such as Bi$_2$Se$_3$ or
Bi$_2$Te$_2$Se, placed on top of an ordinary s-wave superconductor (SC),
subject to applied magnetic field along the axis of the
nanowire. Through a combination of analytical insights and numerical
calculations we demonstrate below that when the magnetic flux through
the nanowire cross-section is close to a half-integer multiple of the
fundamental flux quantum $\Phi_0=hc/e$, the topologically protected
surface state realizes a one-dimensional {\em topological
superconductor} \cite{kitaev1} with Majorana fermions localized near
the ends of the wire. We note that single-crystalline Bi$_2$Se$_3$
nanowires with ribbon geometry (i.e. `nanoribbons') have been
fabricated \cite{peng1} and transport
experiments in these show unambiguous evidence for the topologically
protected surface states up to, possibly, a room temperature
\cite{tang1}. Very recently, the SC proximity effect has been
demonstrated in Sn-Bi$_2$Se$_3$ interfaces \cite{yang1}.

\begin{figure}[b]
\includegraphics[width = 6.0cm]{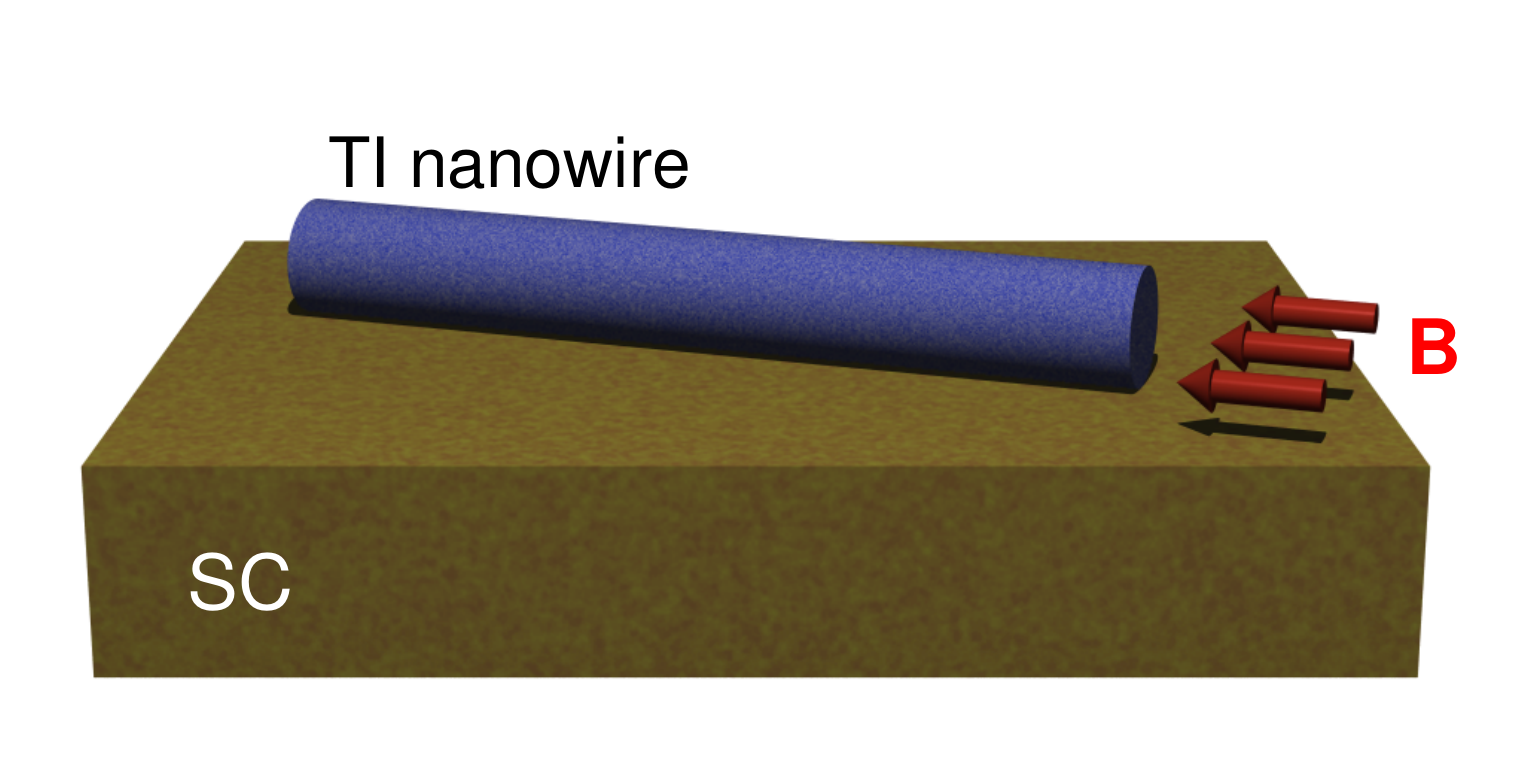}
\caption{Schematic of the proposed device. Magnetic field $\bB$ is applied along the axis of the wire taken to coincide with the $z$-direction.  
}
\end{figure}

The device depicted in Fig.\ 1 appears superficially similar to the
devices based on semiconductor wires
proposed in Refs.\ \cite{sau2,oreg1}. The physics underlying the
emergence of Majorana fermions is nevertheless fundamentally different. In the
semiconductor wires the applied magnetic field serves to open up a gap 
through the
Zeemann coupling to electron spins whereas our proposal relies
exclusively on the
orbital effect of the applied field. This difference leads to several key
advantages of our device over the previous proposals. First, for the
topological phase to occur in a semiconductor wire it is essential that
the chemical potential is fine-tuned to lie inside the Zeemann gap, 
whose typical size in 1T field is $\sim 1$ meV or less. In our
device, by contrast, chemical potential can be anywhere inside the TI
bulk gap, which is $\sim 300$ meV in Bi$_2$Se$_3$. Second, as
explained below, our device can be operated in the regime where the TI 
surface state is protected by time reversal symmetry $({\cal T})$
and the induced pairing gap is therefore robust against  
non-magnetic disorder. Such a
protection is absent in ordinary semiconductor wires \cite{sau2,oreg1}. Finally, the energy gap protecting the Majorana fermions in our setup is generically an order of magnitude larger than in previous proposals.

\begin{figure}[t]
\includegraphics[width =8cm]{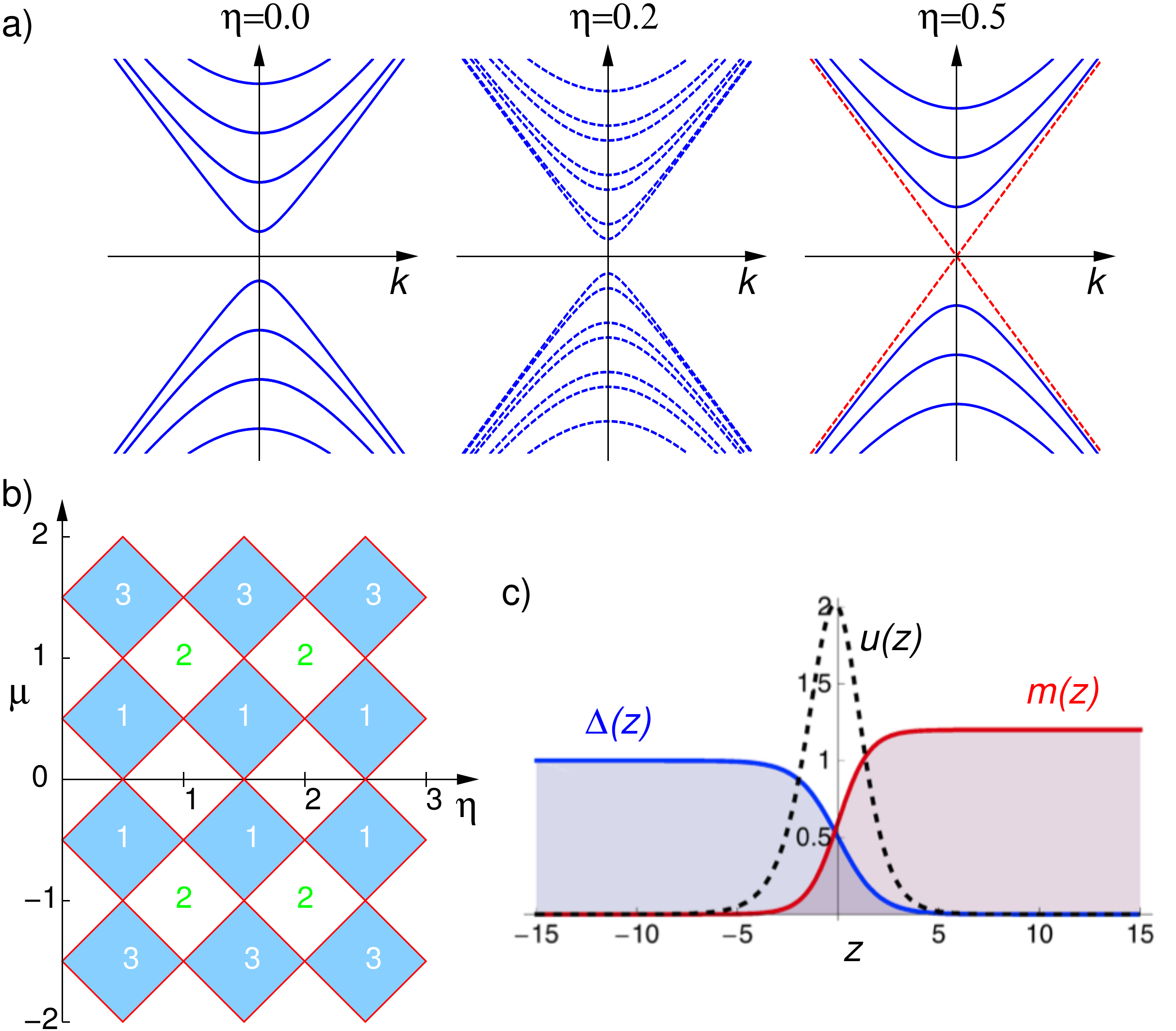}
\caption{a) Surface state excitation spectra $E_{kl}$ for various
  values of magnetic flux $\Phi=\eta\Phi_0$. Solid and dashed lines
  indicate doubly degenerate and non-degenerate bands,
  respectively. b) Kitaev's Majorana number: $\cM=-1 (+1)$ in shaded
  (white) regions. Numerals inside the squares indicate the number of
  Fermi points for $k>0$. c) A possible shape of SC/magnetic domain
  wall located near $z=0$. Dashed line shows the exact zero-mode solution $u(z)$ for this
  domain wall.  
}
\end{figure}
We begin with a qualitative discussion of the physics behind the
proposed device. As explained in previous works \cite{kitaev1,fu2} the
key ingredient necessary to build a topological superconductor is an
underlying normal state characterized by the electron dispersion with
the spin degeneracy removed. It is possible to achieve this situation
on the surface of a TI \cite{mooreN,hasan_rev,qi_rev}. The low-energy fermionic excitations on such a surface are governed by the Dirac
Hamiltonian \cite{mirlin1}
\begin{equation}\label{dir1} h={1\over 2}v\bigr[\hbar\nabla\cdot\bn +
\bn\cdot(\bp\times\bs) + (\bp\times\bs)\cdot\bn\bigl] +\bs\cdot\bmm,
\end{equation}
where $\bp=-i\hbar\nabla$, $v$ is the Dirac velocity, $\bn$ unit vector normal to the surface, and $\bs=(s_1,s_2,s_3)$ the vector
of Pauli spin matrices. $\bmm$ denotes the magnetization vector,
caused e.g.\ by the Zeemann coupling of spins to the external magnetic
field. Inclusion of the latter is not essential for the functionality of the proposed device but will prove useful in subsequent considerations.

Now consider a TI wire in the shape of a cylinder with radius $R$ and magnetic field $\bB$ applied along its axis. 
The magnetic flux is included by replacing the
momentum operator with $\bpi=\bp-(e/c)\bA$, where
$\bA=\eta\Phi_0(\hat{z}\times\br)/2\pi r^2$ is the vector
potential. $\Phi=\eta\Phi_0$ represents the total magnetic flux
through the cylinder. Taking 
$\bn=(\cos{\varphi},\sin{\varphi},0)$ and $\bmm=0$ the spectrum of Hamiltonian
(\ref{dir1}) reads \cite{rosenberg1}
\begin{equation}\label{ekl} 
E_{kl}=\pm v\hbar\sqrt{k^2+{(l+{
1\over2}-\eta)^2\over R^2}}.
\end{equation}
Here $k$ labels momentum eigenstates along the cylinder while
$l=0,\pm1,\dots$ is the angular momentum.
The spectrum in Eq.\ (\ref{ekl}) is clearly periodic in $\eta$ which
reflects the expected $\Phi_0$-periodicity in the total flux. Our
identification of the suitable `spinless' normal state hinges on the
following observation.  For $\eta=0$ all branches of $E_{kl}$ are
doubly degenerate (Fig.\ 2a).  For $\eta\neq 0$, however, the
degeneracy is lifted and one can always find a value of the chemical
potential $\mu$ that yields a {\em single pair} of non-degenerate
Fermi points, as illustrated in Fig.\ 2a. Pairing induced by the
proximity effect in such a state is then expected to drive the system
into a topological phase.

One can formalize the above argument by considering Kitaev's Majorana
number $\cM$ defined as $\cM=(-1)^\nu$, where $\nu$ represents the
number of Fermi points for $k>0$. In the limit of weak pairing,
$\cM=-1$ indicates the existence of unpaired Majorana fermions at the
ends of the wire \cite{kitaev1}.  Fig.\ 2b shows $\cM$ calculated from
the spectrum Eq.\ (\ref{ekl}) as a function of $\mu$ and $\eta$. We
observe, specifically, that when $\eta=1/2$, i.e\ for the flux equal
to half-integer multiple of $\Phi_0$, Majorana fermions will appear
for {\em any} value of the chemical potential (as long as it lies
inside the bulk gap). This result is easily understood by noting that
for $\eta=1/2$ the gapless $l=0$ branch is non-degenerate while the
remaining branches are all doubly degenerate. Thus, the number of
Fermi points for $k>0$ is odd for any value of $\mu$. It is also worth
noting that, for the surface state, $\eta=1/2$ represents a ${\cal
T}$-invariant point and the above pattern of degeneracies should
therefore be robust with respect to non-magnetic disorder 
\cite{bardarson1,yizhang1,potter1}. In this situation Cooper pairs are
formed from time-reversed electron states and the pairing gap is
protected against disorder by Anderson's theorem. Below, we
will explicitly demonstrate the existence and the robustness of the
Majorana fermions both analytically within the low-energy theory based
on Hamiltonian (\ref{dir1}) and numerically using a minimal lattice
model.

Writing the Hamiltonian (\ref{dir1}) in cylindrical coordinates and
with the ansatz for the wavefunction
\begin{equation}\label{psi1} \psi_{kl}(z,\varphi)=e^{i\varphi l}e^{-ikz}\begin{pmatrix} f_{kl} \\
e^{i\varphi}g_{kl} \end{pmatrix} 
\end{equation}
the spinor $\tilde\psi_{kl}=(f_{kl},g_{kl})^T$ is an eigenstate of
\begin{equation}\label{dir2}
\tilde h_{kl}=s_2k+s_3[(l+{1\over2}-\eta)/R+m]. 
\end{equation}
Here we take $v=\hbar=1$ and $\bmm=m\hat{z}$.  To
illustrate the emergence of Majorana fermions in the simplest possible
setting we now focus on the $\eta=1/2$ case and consider chemical
potential $|\mu|< v\hbar/R$, i.e. intersecting only the $l=0$ branch
of the spectrum Eq.\ (\ref{ekl}). The Hamiltonian for this branch then
becomes $h_k=(ks_2-\mu)+ms_3$, where we have explicitly included the
chemical potential term.

With this preparation we can now construct the Bogoliubov-de Gennes
Hamiltonian describing the proximity-induced superconducting order in
the nanowire. In the second-quantized notation it reads
$H=\sum_k\Psi^\dagger_k\cH_k\Psi_k$ with
$\Psi_k=(f_k,g_k,f^\dagger_{-k},g^\dagger_{-k})^T$ and
\begin{equation}\label{hsc1}
\cH_k=\begin{pmatrix} h_k & \Delta_k \\
-\Delta^*_{-k} & -h^*_{-k}
 \end{pmatrix}.
\end{equation}
In the following we consider the simplest $s$-wave pair potential
$\Delta_k=\Delta_0is_2$ with $\Delta_0$ a (complex) constant order
parameter, which corresponds to the pairing term
$\Delta_0(f^\dagger_{k}g^\dagger_{-k}-g^\dagger_{k}f^\dagger_{-k})$. Introducing
Pauli matrices $\tau_\alpha$ in the Nambu space we can write, assuming
$\Delta_0$ real,  
\begin{equation}\label{hsc2}
\cH_k=\tau_3(s_2k-\mu +s_3m)-\tau_2s_2\Delta_0,
\end{equation}
with eigenvalues $E_k=\pm[k^2+\mu^2+m^2+\Delta^2\pm2(k^2\mu^2+\mu^2m^2+m^2\Delta^2)^{1/2}]^{1/2}$. 

In the special case when $\mu=0$ the spectrum simplifies,  
\begin{equation}\label{esc2}
E_k=\pm\sqrt{k^2+(m\pm\Delta_0)^2}.
\end{equation}
The form of the spectrum above suggests that a localized zero-mode
will exist at a boundary between SC and magnetic domains, i.e.\ when
$(m\pm\Delta_0)$ changes sign. We thus seek a zero-energy eigenstate
$\cH\Psi_0(z)=0$ of
$\cH=\tau_3[-s_2i\partial_z+s_3m(z)]-\tau_2s_2\Delta(z)$ with $m(z)$,
$\Delta(z)$ of the form indicated in Fig. 2c. A single Jackiw-Rossi
zero mode \cite{jackiw1} indeed exists and has the form
$\Psi_0(z)=(1,-1,1,-1)^Tu(z)$ with
$u(z)=u_0\exp\int_0^zdz'[\Delta(z')-m(z')]$ and $u_0$ a normalization
constant. The associated field operator
\begin{equation}\label{maj1}
\hat\psi_0=\int u(z) [f(z)-g(z)+f^\dagger(z)-g^\dagger(z)] dz,
\end{equation}
has the property $\hat\psi_0^\dagger=\hat\psi_0$ and represents,
therefore, a Majorana fermion.

The above explicit calculation establishes the existence of an
unpaired Majorana mode at a SC/magnetic domain wall in a TI nanowire
under very special conditions. We now argue that the effect is in fact
generic. First, we reason that the magnetic order, although convenient
in the derivation, is in fact irrelevant. Consider a nanowire of
length $L\gg\xi$, the lengthscale of the zero mode, with the domain
wall located near its center. In a physical system Majoranas always
come in pairs. Since the second Majorana fermion evidently cannot live
in the gapped bulk (or at the magnetic end) we conclude that it must
be localized at the SC end, irrespective of the exact boundary
condition. Second, it is easy to see that the chemical potential can
be moved away from zero without perturbing the Majoranas. Indeed, with
$m=0$ the spectrum of Eq.\ (\ref{hsc2}) reads
$E_k=\pm[(k\pm\mu)^2+\Delta_0^2]^{1/2}$ indicating that the bulk of
the wire shows a SC gap $\Delta_0$ for any value of $\mu$. Therefore,
Majorana end-states will persist even as $\mu$ is varied away from
0. When the chemical potential intersects additional bands then each
band contributes a single Majorana end-state. Any even number of these
will pair up to form ordinary fermions (whose energies will
generically be non-zero) but for an odd number of occupied bands a
single unpaired zero-energy Majorana will remain. This consideration
elucidates the physical meaning of Kitaev's Majorana number $\cM$.  We
note that the topological nature of the zero mode guarantees its stability
against smooth deformations of the nanowire shape, as long as its
bulk remains gapped and the total magnetic flux seen by the surface
state is unchanged.
\begin{figure}[t]
\includegraphics[width = 8.9cm]{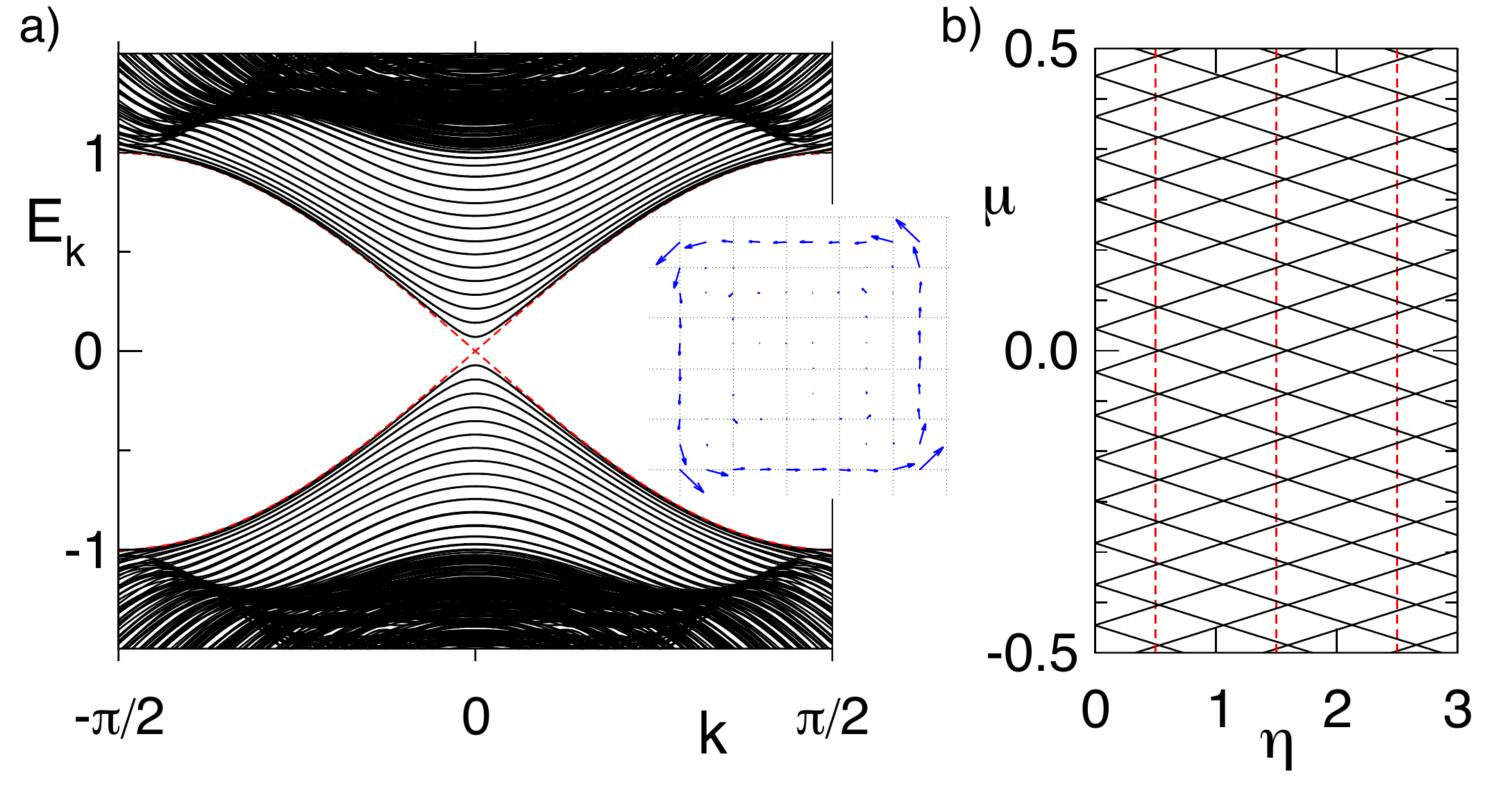}
\caption{a) Energy dispersion for an infinitely long TI wire with $20\times 20$
base described by lattice Hamiltonian (\ref{latt1}) in the normal
state with $\eta=0.52$. For clarity only the low-energy portion of the
spectrum is displayed in a part of the Brillouin zone. Inset shows the
spin expectation values for the gapless state at small positive
$k$. The length of the arrow is proportional to the wavefunction
amplitude. b) Lines separating regions with different Majorana number
$\cM=-1 (+1)$ extracted from the spectrum. All energies are in units
of $\lambda=150$meV and we use parameters $\lambda_z=t=1$,
$\epsilon=4$ and $g=32$, corresponding to the strong TI phase with
Z$_2$ index (1;000) and bulk bandgap $2\lambda=300$meV.
}
\end{figure}

To explicitly address the existence and robustness of Majorana
end-states we now study
the nanowire using a concrete lattice model of Bi$_2$Se$_3$ family
of materials \cite{qi_rev}. Specifically, we use the model given by 
Fu and Berg \cite{fu-berg1} regularized on a simple cubic lattice, defined by a
$k$-space Hamiltonian
\begin{equation}\label{latt1} 
h_\bk=M_\bk\sigma_1+\lambda\sigma_3(s_2\sin{k_x}-s_1\sin{k_y})+\lambda_z\sigma_2\sin{k_z},
\end{equation}
with $M_\bk=\epsilon-2t\sum_\alpha\cos{k_\alpha}$.  Here $\sigma_\alpha$
represent the Pauli matrices acting in the space of two
independent orbitals per lattice site. For $\lambda, \lambda_z>0$ and
$2t<\epsilon<6t$ the system described by Hamiltonian (\ref{latt1}) is
a TI in Z$_2$ class (1;000), i.e.\ a strong topological insulator. The
 magnetic field enters through the Peierls
substitution, replacing all hopping amplitudes as $t_{ij}\to
t_{ij}\exp{[-(2\pi i/\Phi_0)\int_i^j\bA\cdot d{\bf l}]}$ and the Zeemann
term $-g\mu_B \bB\cdot{\bs}/2$ where $\mu_B=e\hbar/2m_ec$ is the Bohr
magneton. In the SC state the BdG Hamiltonian takes the form of Eq.\
(\ref{hsc1}) with  $\Delta_\bk=\Delta_0is_2$ describing on-site
spin singlet pairing.

We have solved the problem posed by Hamiltonian (\ref{latt1}) in
various wire geometries by exact numerical diagonalization and by
sparse matrix techniques. Fig.\ 3a shows a typical example of the
excitation spectrum in an infinitely long wire with $W\times W$
cross-section in the normal state. We observe that for $\eta$ close to
$1/2$ the surface state is indeed gapless and the low-energy modes
exhibit the expected pattern of degeneracy. Because of the surface
state penetration into the TI bulk the surface electrons see a
slightly smaller magnetic flux than the nominal flux $\Phi=BW^2$
given by the wire geometry and the gapless state is shifted to a
slightly higher value of $\eta$. This is also seen in Fig. 3b which
displays the Majorana number for the same system. This figure
indicates that for $\eta=0.52(2n+1)$ the system will be a 1D
topological SC for any value of $\mu$ inside the bulk gap.

\begin{figure}[t]
\includegraphics[width = 8.9cm]{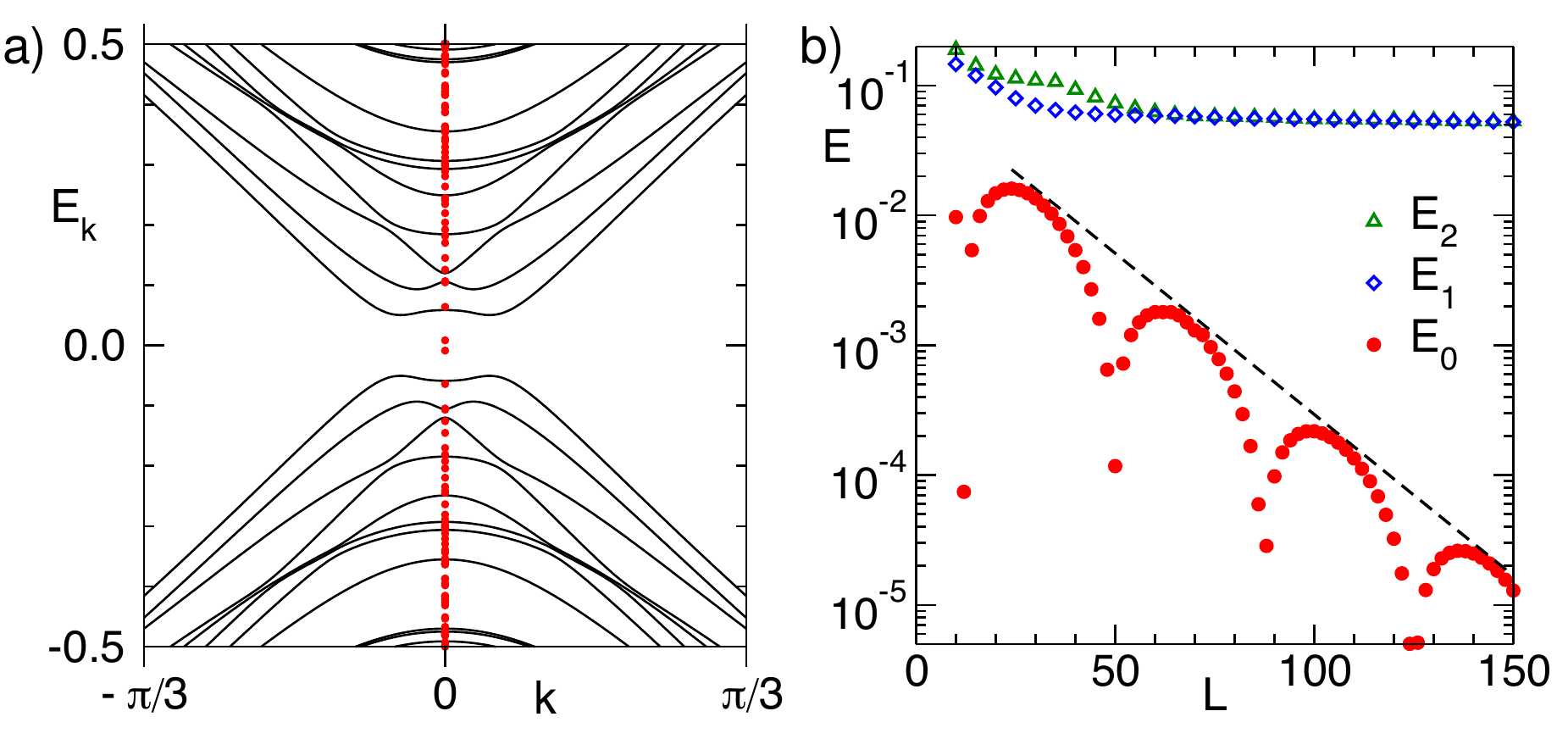}
\caption{a) Energy bands for an infinitely long TI wire with
  $6\times 6$ base in the  SC  state with $\eta=0.49$, $\mu=0.09$,
  $\Delta_0=0.08$ and $g=0$ (solid lines), and  the energy levels for
  $L=36$ finite-length wire with open boundary conditions (red circles) obtained by exact
  numerical diagonalization.
 b) Three lowest positive energy eigenvalues obtained by the Lanczos
 method as a function of
 $L$. Dashed line represents the envelope function $0.089e^{-L/\xi}$
 with $\xi=17.5$.
}
\end{figure}
Superconducting order opens a gap in the electron excitation spectrum
as illustrated in Fig.\ 4a. For an open-ended wire, crucially, our
calculations reveal a pair of non-degenerate states at $\pm E_0$ inside the SC
gap whose energies approach zero for large $L$ as
$E_0\propto e^{-L/\xi}$. Fig.\ 4b illustrates this exponential decay
(which is in addition modulated by oscillations at $2k_F$). Higher energy
eigenstates approach non-zero values close to $\Delta_0$ for large
$L$. We have verified that the appropriate linear superpositions of
the wavefunctions associated with the $\pm E_0$ eigenvalues are exponentially 
localized near the ends of the wire. The
corresponding field operators then satisfy the Majorana
condition $\hat{\psi}^\dagger=\hat{\psi}$, and represent, up to
exponentially small corrections in their separation, the  
Majorana zero modes.

We conclude with comments on the experimental realization. For the
existing Bi$_2$Se$_3$ nanowires \cite{peng1,tang1} with cross-sectional
area $S\approx 6\times 10^{-15}$m$^2$ the surface level spacing is $\delta
E_S\simeq v\hbar\sqrt{\pi/S}\simeq 7$meV. At half flux quantum, which
corresponds to the  magnetic field strength $B=\Phi_0/2S\simeq 0.34$T,
the Zeemann
energy scale $\delta E_Z=g\pi\hbar^2/2m_eS\simeq 0.6$meV (taking
$v=5\times 10^{5}$m/s and $g=32$) and thus probably negligible. 
Experiments on planar Sn-Bi$_2$Se$_3$ 
interfaces \cite{yang1} show induced SC gap $\sim 0.2$meV, a significant
fraction of the native Sn bulk gap ($\sim 0.6$meV). It thus appears
conceivable that a pairing gap of several meV could be induced in
Bi$_2$Se$_3$ nanowires by using in place of Sn a superconductor with 
larger bulk gap, such as NbTiN or MgB$_2$. A gap of this size should
permit detection of the Majorana fermion by scanning tunnelling
techniques. The latter will show up as a zero-bias peak exponentially
localized near the end of the wire in the topological phase, but will
disappear as the phase boundary into non-topological phase is traversed
by tuning either chemical potential or magnetic field.  Unambiguous
detection of Majorana fermions can be achieved by probing
$4\pi$-periodic Josephson current through a weak link in the wire as
described in Refs.\   
\cite{sau2,oreg1,kitaev1}. Another intriguing possibility is to
fabricate nanowires from Cu$_x$Bi$_2$Se$_3$ which becomes a
superconductor below 4K \cite{ong1}
while simultaneously retaining the protected surface states
 \cite{wray1}. 

Energy scales order of magnitude higher compared to the ordinary
semiconductor wires, significantly reduced requirements for the
chemical potential control and the sample purity (afforded by
operation in the ${\cal T}$-invariant regime) make TI
nanowires promising candidates for the first experimental detection
of Majorana
fermions. Our numerical simulations of the model Hamiltonian (\ref{latt1}) with disorder (to be reported separately) show essential robustness against non-magnetic impurities \cite{ashley2} but systematic studies of the
effects of disorder and interactions along the lines of recent works
\cite{brouwer1,fisher1} constitute an exciting future research direction.

The authors are indebted to J.\ Alicea, P.\ Brouwer, S.\ Frolov, L.\
Fu, I.\ Garate, G. Refael, and X.-L. Qi for valuable comments and
discussions. The work was supported by NSERC and CIfAR.

\end{document}